\documentclass[prd,aps,twocolumn,nofootinbib,preprintnumbers,superscriptaddress,preprintnumbers,balancelastpage,longbibliography]{revtex4-2}
\usepackage[english]{babel}

\usepackage{amsmath,mathtools,physics,xfrac}
\usepackage[normalem]{ulem}
\usepackage{graphicx}
\usepackage{afterpage}
\usepackage{float}
\usepackage{subfigure}
\usepackage{rotating}
\usepackage{multirow}
\usepackage{tabularx}
\usepackage{booktabs}
\usepackage{fancyhdr}
\usepackage[utf8]{inputenc}
\usepackage{theorem}
\usepackage{moreverb}
\usepackage{euscript}
\usepackage{psfrag}
\usepackage{slashed}
\usepackage{mathtools}
\usepackage{makecell}
\usepackage{adjustbox}
\usepackage{dcolumn}
\usepackage{bm}
\usepackage[dvipsnames]{xcolor}
\usepackage{graphics}
\usepackage{hyperref}
\usepackage{lettrine}
\usepackage{amsmath}

\newcommand{\refjnl}[1]{{\rm#1}}

\def\apj{\refjnl{ApJ}}                 
\def\apjl{\refjnl{ApJ}}                
\def\aap{\refjnl{A\&A}}                

\def\mnras{\refjnl{MNRAS}}             
\def\ssr{\refjnl{Space~Sci.~Rev.}}     

\input Zallman.fd

\LettrineTextFont{\itshape}
\usepackage{tikz}
\usetikzlibrary{arrows.meta,positioning,calc,shapes,fit,decorations.text}

\newcommand{\Ha}{\mathrm{H}\alpha}

\definecolor{halpha}{HTML}{FF0000}

\hypersetup{
     colorlinks   = true,
     citecolor    = halpha,
     urlcolor     = halpha,
     linkcolor    = halpha
}

\usepackage{color,xcolor}

\definecolor{darkblue}{rgb}{0.0,0.0,0.75}
\definecolor{darkred}{rgb}{0.6,0.0,0}
\definecolor{darkgreen}{rgb}{0.0,0.6,0.}

\newcommand\redsout{\bgroup\markoverwith{\textcolor{red}{\rule[0.5ex]{2pt}{0.4pt}}}\ULon}

\begin{document}

\preprint{SLAC-PUB-251031}

\title{Search for Dark Matter Annihilation and Decay with H$\alpha$ Line Emission}

\author{Rebecca~K.~Leane}
\thanks{\href{mailto:rleane@slac.stanford.edu}{rleane@slac.stanford.edu}; \href{http://orcid.org/0000-0002-1287-8780}{0000-0002-1287-8780}
}
\affiliation{SLAC National Accelerator Laboratory, 2575 Sand Hill Rd, Menlo Park, CA 94025, USA}
\affiliation{Kavli Institute for Particle Astrophysics and Cosmology, Stanford University, Stanford, CA 94305, USA}

\date{\today}

\begin{abstract}

I present a new indirect search for dark matter (DM) using Hydrogen-$\alpha$ (H$\alpha$) recombination emission. DM annihilation or decay products can ionize neutral gas; subsequent recombination cascades generate H$\alpha$ photons through the $3\rightarrow2$ transition. In quiet gas-rich dwarf galaxies, the $n{=}2$ population is negligible, so H$\alpha$ is effectively unabsorbed and traces the DM-energy injection site.  Using the non-detection of extended H$\alpha$ emission in the Leo T dwarf galaxy with Multi Unit Spectroscopic Explorer (MUSE) observations, I derive the first H$\alpha$-based limits on DM annihilation and decay, reaching leading sensitivity for parts of the eV--GeV mass range. Existing and upcoming telescopes can further extend this reach, establishing H$\alpha$ imaging as a powerful DM search strategy.
\end{abstract}

\maketitle

\lettrine{T}{he hydrogen atom} is a curious one. Despite containing only one electron and proton, it shines and hides with a vast forest of different emission and absorption lines, organized into series of bound-bound transitions. The first discovered, the Balmer series, collects all transitions to $n{=}2$ and sits mostly in the optical. Its brightest line, H$\alpha$ ($n=3{\rightarrow}2$), gives many emission nebulae their deep red glow, perhaps most recognizably outlining the Horsehead Nebula.

H$\alpha$ is both ubiquitous and observationally convenient. Its red shine suffers relatively little dust extinction, and in ionized gas a substantial fraction of recombination cascades pass through the $3{\to}2$ transition, so H$\alpha$ closely tracks the local recombination rate. Consequently, H$\alpha$ imaging and spectroscopy are standard probes of ionization from massive-star photoionization, shocks, and cosmic rays~\cite{RydenPogge2021_IISM}. This makes H$\alpha$ an efficient tracer for processes that inject ionizing power into neutral gas.

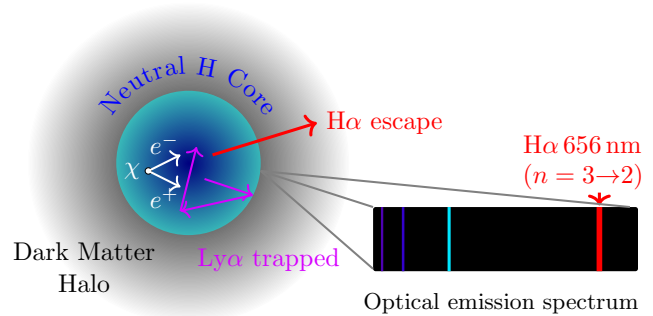
\begin{figure}[t]
\centering
\scalebox{0.97}{\definecolor{brightpurp}{RGB}{214,0,255}
\begin{tikzpicture}[font=\normalsize, scale=1.1]

\shade[shading=radial, inner color=black, outer color=white, draw=none]
       (-7,1.2) circle (2);
 \node at (-8.3,0.1) [black] {Dark Matter};
 \node at (-8.3,-0.3) [black] {Halo};

\shade[shading=radial, inner color=blue, outer color=cyan, opacity=0.5, draw=none]
      (-7,1.2) circle (0.9);

\path[
  decorate,
  decoration={
    text along path,
    text={|\color{blue}\normalsize|Neutral H Core},
    text align={align=center},
    raise=0.5ex 
  }
] (-7,1.2) ++(150:1.05) arc[start angle=150, end angle=30, radius=1.05];

\draw[fill=white] (-7.5,1.1) circle (0.04);
\draw[->,thick,white] (-7.5,1.1) -- (-7.1,1.3) node[midway, above] {$e^-$};
\draw[->,thick,white] (-7.5,1.1) -- (-7.1,0.9) node[midway, below] {$e^+$};
\node at (-7.7,1.1) [white] {$\chi$};

\draw[->,very thick,red] (-6.7,1.3) -- (-5.4,1.7) node[right, red] {H$\alpha$ escape};
\draw[->,thick,brightpurp] (-6.8,1.0) -- (-6.2,0.8);
\draw[->,thick,brightpurp] (-6.2,0.8) -- (-7.1,0.6);
\draw[->,thick,brightpurp] (-7.1,0.6) -- (-6.9,1.4);
\node at (-6,0) [brightpurp] {Ly$\alpha$ trapped};

\begin{scope}[shift={(-4.7,-0.15)}]
  \def\Xscale{0.011}      
  \def\ybase{0}
  \def\height{0.8}
  
  \fill[black, rounded corners=1pt]
       (0,\ybase) rectangle ({\Xscale*(700-400)},{\ybase+\height});

\foreach \lam/\col/\lw in {
  410.2/blue!40!violet/0.9, 
  434.0/blue!60!violet/1.0, 
  486.1/blue!10!cyan/1.2,   
  656.3/red/2.2             
}{
  \draw[line width=\lw pt, \col]
    ({\Xscale*(\lam-400)},\ybase) -- ({\Xscale*(\lam-400)},{\ybase+\height});
}

  \draw[->, very thick, red]
    ({\Xscale*(656.3-400)}, {\ybase+\height+0.18})
      -- ({\Xscale*(656.3-400)}, {\ybase+\height+0.02});
\node[red, right]
  at ({\Xscale*(155) + 0.06}, {\ybase+\height+0.78})
  {H$\alpha$\,656\,nm };
\node[red, right]
  at ({\Xscale*(160) + 0.0}, {\ybase+\height+0.38})
  {($n=3{\to}2$)};

   \node[align=left] at ({\Xscale*(560-415)}, {\ybase+\height-1.2})
     {\small Optical emission spectrum};
 \end{scope}
 
\coordinate (CorePt) at (-6.1,1.1);

\coordinate (SpecTL) at (-4.7,0.65);
\coordinate (SpecBL) at (-4.7,-0.12);
\coordinate (SpecTR) at (-1.5,0.65);

 \draw[thick,gray] (CorePt) -- (SpecTL);
 \draw[thick,gray] (CorePt) -- (SpecBL);
 \draw[thick,gray] (CorePt) -- (SpecTR);

\end{tikzpicture}}
\caption{Schematic of H$\alpha$ emission for DM discovery in quiet gas-rich galaxies. DM annihilation or decay products ionize neutral hydrogen, which then recombines to emit $\Ha$ radiation that is detectable with optical telescopes.}\label{fig:schematic}
\end{figure}

In this \textit{Letter}, I propose using the H$\alpha$ line to search for dark matter (DM). DM annihilation or decay to Standard Model (SM) particles can ionize neutral hydrogen (H\,\textsc{i}); subsequent recombination cascades generate detectable H$\alpha$ photons. In quiet, gas-rich dwarf galaxies where ionized (H\,\textsc{ii}) regions are absent, the $n{=}2$ population is tiny, making H$\alpha$ effectively non-resonant and unabsorbed; the resulting emission traces the DM energy-injection site. The idea is schematically summarized in Fig.~\ref{fig:schematic}.

H$\alpha$ as a DM signal offers several advantages. First, it can be used to probe a wide range of DM masses, down to hydrogen's ionization threshold, below the reach of many conventional indirect searches~\cite{Colafrancesco:2006he,Baltz:2008wd,Jeltema:2008ax,Atwood_2009,Bergstrom:2013jra,Elor:2015bho,Fermi-LAT:2015ycq,Fermi-LAT:2015att,Cavasonza:2016qem,Cuoco:2016eej,Cui:2016ppb,Cuoco:2017rxb,Regis:2017oet,Slatyer:2017sev,Leane:2018kjk,Hooper:2019xss,Alvarez:2020cmw,
Leane:2020liq,McDaniel:2023bju}. This allows a new probe of sub-GeV DM, which is well motivated and appears in a broad class of models, including thermal light-dark-sector scenarios, freeze-in DM, and bosonic candidates such as axion-like particles, dark photons, and sterile neutrinos; see, $e.g.$, Refs.~\cite{Adams:2022pbo,Boyarsky:2018tvu,Essig:2022dfa}. Second, in terms of ease of discovery, the broad applicability of existing $\Ha$ architecture and astrophysical searches means that data largely does not need to be reprocessed or tailored to specific DM masses: the smoking gun is simply the 656 nanometer $\Ha$ line. This complements conventional line searches that aim to detect the annihilation/decay products at or near the DM mass~\cite{Essig:2013goa,Fermi-LAT:2013thd,Ibarra:2015tya,HESS:2018cbt,MAGIC:2022acl,Bulbul:2014sua,Boyarsky:2014jta,PhysRevLett.120.061301,Regis:2020fhw,Yin:2024lla,Bessho:2022yyu,Janish:2023kvi,Roy:2023omw,Pinetti:2025owq,Saha:2025any}. Third, it is a present-day probe of DM in nearby systems, providing complementarity on systematics for early-Universe or high-redshift DM-energy injection searches~\cite{Adams:1998nr,Chen:2003gz,Padmanabhan:2005es,Zhang:2007zzh,Galli:2009zc,Slatyer:2009yq,Kanzaki:2009hf,Hisano:2011dc,Hutsi:2011vx,Galli:2011rz,Finkbeiner:2011dx,Pritchard:2011xb,Slatyer:2012yq,Diamanti:2013bia,Galli:2013dna,Madhavacheril:2013cna,Slatyer:2015jla,Slatyer:2016qyl,Poulin:2016anj,Liu:2016cnk,Planck:2018vyg, Acharya:2019uba,Liu:2020wqz,Qin:2023kkk,Agius:2025nfz}. For other types of present-day astrophysical DM-ionization signals, see Refs.~\cite{Sciama1990ApJ,1998ApJ...505L..35S,Prabhu:2022dtm,Blanco:2023qgi,Blanco:2023bgz,Leane:2024bvh,Blanco:2024lqw,Blanco:2025wpo}. 

In the following, I review H$\alpha$ production in neutral gas and derive the DM-induced H$\alpha$ line signal for DM annihilation or decay to photons and electrons. I apply this framework to Leo~T, a nearby ($\sim\!420$\,kpc) DM-dominated dwarf galaxy whose central H\,\textsc{i} core and negligible current star formation minimize conventional H$\alpha$ backgrounds. Using recent observations with the Multi Unit Spectroscopic Explorer (MUSE) on the Very Large Telescope (VLT)~\cite{Vaz_2023}, I obtain the first H$\alpha$ DM limits.  Finally, I conclude and discuss the prospects for further developments.\\


\noindent\textbf{\textit{Overview of $\Ha$ Emission in Neutral Gas.---}} I target the neutral gas in the central regions of gas-rich dwarfs. For Leo T, the inner core contains both a cold and a warm neutral medium (WNM)~\cite{Ryan-Weber:2007guk,Adams_2018,Faerman:2013pmm,Patra_2018,Vaz_2023}; I conservatively restrict to the WNM, which is better characterized. The gas is H dominated with primordial He and negligible metals~\cite{Simon:2007dq,Ryan-Weber:2007guk,Vaz_2023}. I do not consider the outer star-formation-correlated ionized regions as signal regions, due to higher backgrounds.

Ionizing sources can inject energetic electrons and/or photons into H\,\textsc{i}. As fast electrons slow, they ionize additional H atoms, excite H without ionizing it, and lose sub-threshold energy as heat~\cite{RydenPogge2021_IISM}. In quiescent gas, ionizations are usually quickly balanced by recombinations, so the emergent $\Ha$ is set by the fraction of deposited power that creates ion-electron pairs, and by the ensuing radiative cascade when those pairs recombine.

Recombinations in neutral gas do not usually place the electron directly into the ground state. Instead, radiative recombination predominantly populates excited bound states ($n\geq2$); the atom then de-excites via a radiative cascade emitting line photons. Because Lyman-series photons are resonant with the ground state, Lyman-continuum and Ly$\alpha$ are trapped and locally reprocessed, in a scenario called ``Case-B recombination"~\cite{RydenPogge2021_IISM}. The escaping cascade thus proceeds from $n\ge2$, with a well-known fraction of recombinations, $f_{\Ha}\simeq0.46$,  passing through $3{\to}2$ and producing $\Ha$, which has energy $E_{\Ha}=1.89$~eV. Unlike Ly$\alpha$, $\Ha$ is non-resonant; with negligible $n{=}2$ populations in cool/warm neutral gas, the H\,\textsc{i} optical depth at H$\alpha$ remains tiny even at large columns. $\Ha$ photons therefore escape essentially unattenuated and preserve the intrinsic morphology, making diffuse $\Ha$ flux a low-systematics measurement of the ionizing power deposited into the neutral phase.\\


\noindent\textbf{\textit{Dark Matter H$\alpha$ Line Signal.---}} Gas-rich dwarf galaxies such as Leo~T host dense DM halos~\cite{Simon:2007dq,Ryan-Weber:2007guk,Adams_2018,Faerman:2013pmm,Patra_2018,Vaz_2023}. DM co-located with the WNM injects energetic particles that ionize H\,\textsc{i}; subsequent recombinations produce escaping H$\alpha$ photons. For a sky region \(A\) (the observing aperture), the predicted H$\alpha$ flux for DM of mass $m_\chi$ annihilating with rate $\langle\sigma v\rangle$ or decaying with lifetime $\tau_\chi$ is
\begin{equation}
F_{\Ha}^{\rm DM}(A)
=\frac{1}{4\pi}\;f_{\rm dep}^{\Ha}(E)\;f_{\rm eq}\;\mathcal{Q}[A],
\label{eq:DMflux}
\end{equation}
with the DM energy-injection flux
\begin{equation}
\mathcal{Q}[A]\equiv
\begin{cases}
\dfrac{\langle\sigma v\rangle}{2\,m_\chi^2}\;J[A]\; E_{\rm ann}, & \text{(annihilation)}\\[8pt]
\dfrac{1}{m_\chi\,\tau_\chi}\;D[A]\; E_{\rm dec}, & \text{(decay)}
\end{cases}
\label{eq:Q_region}
\end{equation}
where the standard $J$ and $D$ factors are
\begin{equation}
J[A]\equiv \int_A d\Omega\!\int_{\rm los} d\ell\,\rho_\chi^2,\ \ \
D[A]\equiv \int_A d\Omega\!\int_{\rm los} d\ell\,\rho_\chi\,,
\label{eq:JD_region_defs}
\end{equation}
where $d\Omega$ is the differential solid angle, $\ell$ is the line-of-sight distance, and $\rho_\chi$ is the DM density. Above, the injected SM energies are
\(E_{\rm ann,\gamma}=2\,m_\chi\), \(E_{\rm dec,\gamma}=m_\chi\) for photons; and
\(E_{\rm ann,e}=2\,(m_\chi-m_e)\), \(E_{\rm dec,e}=m_\chi-2\,m_e\) for electrons; the per-particle energy is half these values.
The factor \(f_{\rm dep}^{\Ha}(E)\) is the fraction of injected SM energy that ultimately emerges in H$\alpha$, and is derived shortly.

The DM-powered ionization and hydrogen recombination reach equilibrium across the bulk of the considered parameter space, such that generally the equilibrium factor $f_{\rm eq}\sim1$. However, at sufficiently low ionization rates, a steady state is not reached. In that case, the H$\alpha$ yield is suppressed and $f_{\rm eq}<1$; see the Supplemental Material for more details.\\

\noindent\textbf{\textit{Dark Matter Annihilation or Decay into Photons.---}} If DM injects photons into the medium, they can be absorbed by the neutral medium, with probability
\begin{equation}
    P_{\rm abs}(E_\gamma)=1-e^{-\tau(E_\gamma)}\,,
\end{equation}
where for photons of energy $E_\gamma$ the optical depth is
\begin{equation}
\tau(E_\gamma)
= \rho_{\rm gas}\, \ell_{\rm gas}
\sum_{i\in\{\mathrm{H},\mathrm{He}\}}
w_i\!\left(\frac{\mu_{\rm en}}{\rho}\right)_{i}(E_\gamma)\,,
\end{equation}
where $w_{\rm H}$ and $w_{\rm He}$ are the mass fractions of H and He respectively, and $\mu_{\rm en}/\rho$ is the mass energy-absorption coefficient for the given element. For $E_\gamma<30$ eV, these are obtained using the photoionization cross sections from Ref.~\cite{1990A&A...237..267B}. For $30<E_\gamma<1000$~eV I use the data from Ref.~\cite{CXRO:atten2}, and for above 1 keV I use Ref.~\cite{HubbellSeltzer_XAAMDI_2004}. The average density of the in-aperture gas is $\rho_{\rm gas}$, and the path length through the gas is $\ell_{\rm gas}$; see the Supplemental Material for system-/telescope-specific values. The fraction of energy absorption by each species is then
\begin{equation}
f_i(E_\gamma)=
\frac{w_i\,(\mu_{\rm en}/\rho)_i(E_\gamma)}%
{\sum_{j\in\{\mathrm{H},\mathrm{He}\}} w_j\,(\mu_{\rm en}/\rho)_j(E_\gamma)},
\ i\in\{\mathrm{H},\mathrm{He}\},
\end{equation}
so that $f_{\rm He}=1-f_{\rm H}$. Assuming the photons are above the relevant ionization thresholds, then the $\Ha$ deposition fraction per photon is
\begin{align}
\label{eq:fdep_gamma_Ha}
f_{\rm dep,\gamma}^{\Ha}(E_\gamma)
&= P_{\rm abs}(E_\gamma)\,\frac{f_{\Ha}E_{\Ha}}{E_\gamma}\Bigg[
  f_{\rm H}(E_\gamma)+f_{\rm He}(E_\gamma)
\\
&+\sum_{i\in\{\mathrm{H},\mathrm{He}\}}
   f_i(E_\gamma)\,
   (E_\gamma-I_i)\,
   \frac{f_{\rm ion}^{\rm H}(E_\gamma-I_i)}{I_\textrm{H}}\,
\Bigg]\,,\nonumber
\end{align}
where $I_{\rm H}=13.6\,$eV and $I_{\rm He}=24.6\,$eV are the ionization energies of hydrogen and helium respectively; for $E_\gamma<I_{\rm He}$ only the H terms contribute. In Eq.~\eqref{eq:fdep_gamma_Ha}, the $f_{\rm H}$ term describes primary H
photoionizations, corresponding to the $\Ha$ produced when a single absorbed photon ionizes hydrogen, even if the photoelectron has negligible kinetic energy. The $f_{\rm He}$ term corresponds to one secondary H photoionization per primary He photoionization~\cite{RydenPogge2021_IISM}, which occurs when He recombines, emits photons below its own ionization threshold, and photoionizes H. The final sum describes additional H ionizations from secondary electrons:
$f_{\rm ion}^{\rm H}$ is the fraction of the secondary-electron energy that goes into H ionizations in a mixed H/He gas as a function of the electron energy~\cite{Furlanetto_2010}, so energy initially absorbed by either H or He can yield further H ionizations. \\


\noindent\textbf{\textit{Dark Matter Annihilation or Decay into Electrons.---}}
When DM injects $e^\pm$ into a predominantly neutral medium, the H$\alpha$ yield is set by turbulent transport, collisional losses, and radiative losses. Because the telescope aperture can be smaller than the WNM core, electrons injected outside the aperture can still diffuse in and deposit energy. I therefore divide the WNM into the aperture region $A$ and its complement $B$, and label the exterior region $C$.

Electrons random-walk on turbulence characterized by an Alfv\'en speed $v_A$ and magnetic-turbulence coherence length $L_{\rm coh}$, defining a diffusion coefficient $K$~\cite{1966ApJ...146..480J,2002cra..book.....S,2022SSRv..218...33E},
\begin{align}
v_A(B,n_{\rm H}) &= \frac{B_{\rm mag}}{\sqrt{4\pi\,\rho_{\rm ion}}}\,,\qquad
K = \frac{1}{3}\,v_A\,L_{\rm coh}\,,
\label{eq:diffusion}
\end{align}
where $B_{\rm mag}$ is the magnetic field, and $\rho_{\rm ion}$ is the ion density. The escape rate for diffusion out of a given zone $X$ into a neighboring zone $Y$ is approximated as~\cite{Redner_2023}
\begin{align}
\Gamma_{X\to Y} ^{\rm esc} &= \frac{\pi^2 K}{\ell_{X\rightarrow Y}^2},\
\end{align}
where ${\ell_{X\rightarrow Y}}$ is the distance to the region boundary. For parameter inputs relevant to specific systems and telescopes, see the Supplemental Material. In each zone, the deposited (collisional) and radiative loss rates are
\begin{align}
\Gamma_X^{\rm dep}(E_e)
&= \frac{c}{E_e}\sum_{i\in\{\mathrm{H},\mathrm{He}\}} m_i\,n_{i,X}\,S_i^{\rm coll}(E_e),
\\
\Gamma_X^{\rm rad}(E_e)
&= \frac{c}{E_e}\sum_{i\in\{\mathrm{H},\mathrm{He}\}} m_i\,n_{i,X}\,S_i^{\rm rad}(E_e)
+ \Gamma_X^{\rm IC+syn}(E_e).\nonumber
\end{align}
where $S_i$ are stopping powers~\cite{NIST_SRD124_Landing}, $m_i$ and $n_i$ are the mass and number density of target $i$ respectively, $c$ is the speed of light, and $\Gamma^{\rm IC+syn}$ accounts for inverse-Compton and synchrotron losses (though this is negligible in my parameter space). The total energy loss is $\Gamma^{\rm tot}_X(E_e)=\Gamma_X^{\rm dep}(E_e)+\Gamma_X^{\rm rad}(E_e)$; note that only the collisional energy loss is converted into the $\Ha$ signal, the radiative losses instead stream energy away.

The inverse mean residence time $\Lambda_B(E_e)$ in zone $B$ is obtained by summing the competing rates for (i) in-situ losses and (ii) diffusive crossings to the neighboring regions,
\begin{equation}
\Lambda_B(E_e)=\Gamma_B^{\rm tot}(E_e)+\Gamma_{B\to A} ^{\rm esc}+\Gamma_{B\to C} ^{\rm esc}\,,
\end{equation}
and the fraction of electrons in $B$ traveling into $A$ is
\begin{equation}
r_B(E_e)=\frac{\Gamma_{B\to A} ^{\rm esc}}{\Lambda_B(E_e)}\,.
\end{equation}
The probability to deposit in $A$ after starting in $A$ is
\begin{equation}
P_{A\to A}(E_e)=
\frac{\Gamma_A^{\rm dep}(E_e)}{\Gamma_A^{\rm tot}(E_e)+\Gamma_{A\to B} ^{\rm esc}\big[1-r_B(E_e)\big]}\,,
\end{equation}
where the second term in the denominator accounts for electrons leaving zone $A$ into $B$; only a fraction $1-r_B(E_e)$ of these are permanently lost to regions outside the aperture; escape directly from $A$ to $C$ is negligible.

In my two-zone transport model, electrons produced anywhere in the WNM core can random-walk into the aperture region $A$ and deposit there. Accordingly, for electrons I replace the $J$ and $D$ factors over the aperture by those over the full WNM core,
\begin{equation}
    \widetilde J = J[A\cup B],\ \ \
\widetilde D = D[A\cup B],
\end{equation}
and in Eq.~\eqref{eq:Q_region} I take $J[A]\to \widetilde J$ and $D[A]\to \widetilde D$ for the electron case. The fraction of the injected power that is ultimately collisionally deposited within the aperture is encoded in $F_A(E_e)$,
\begin{equation}
F_A(E_e)=
f_A\,P_{A\to A}(E_e)+
(1-f_A)\,r_B(E_e)\,P_{A\to A}(E_e)\,,
\label{eq:FA_main}
\end{equation}
where $f_A$ is the DM annihilation or decay-induced birth fraction of $e^\pm$ in $A$ (and $1-f_A$ in $B$), compared to the whole WNM core.

The electron H$\alpha$ deposition fraction is therefore
\begin{equation}
f_{{\rm dep},e}^{\rm H\alpha}(E_e)
=\frac{f_{\rm ion}^H (E_e)}{I_{\rm H}}f_{\rm H\alpha}E_{\rm H\alpha}\;F_A(E_e)\,.
\label{eq:fdepe_mono}
\end{equation}

\begin{figure*}[t]
    \centering
\includegraphics[width=0.49\textwidth]{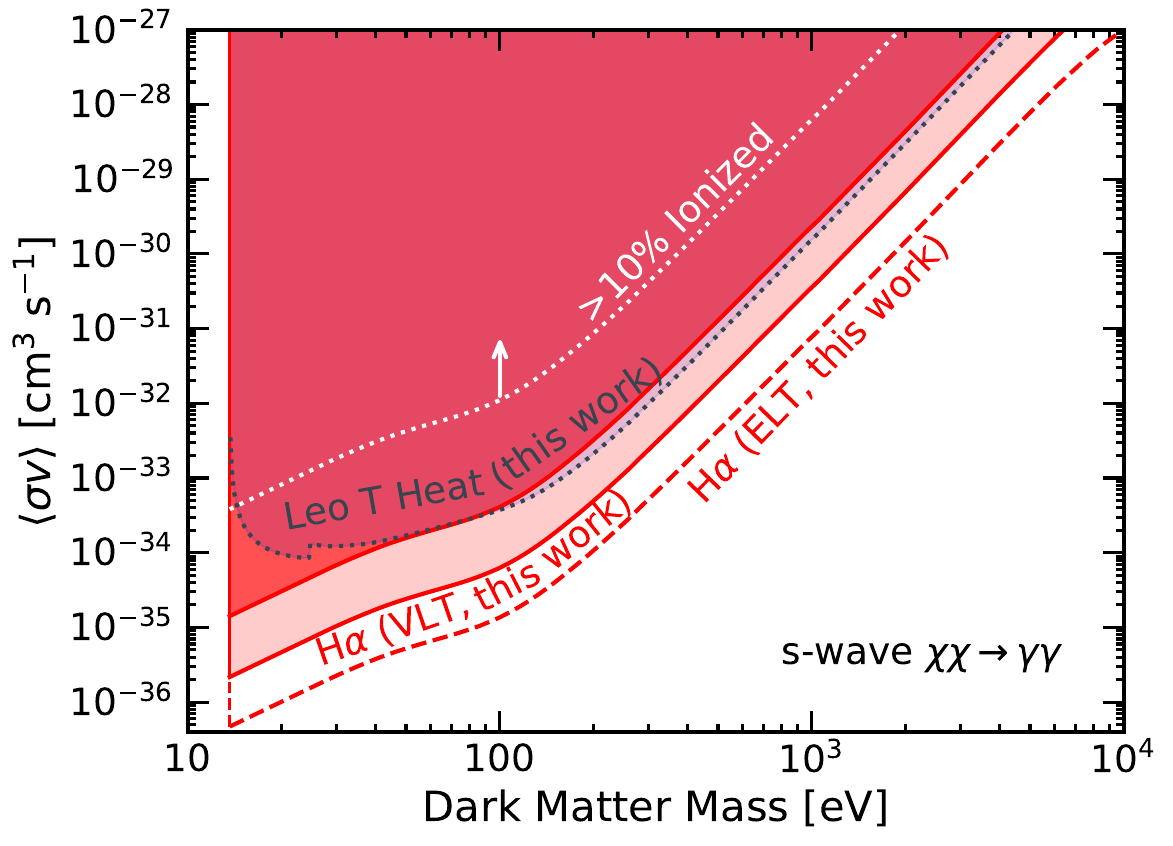}
\includegraphics[width=0.49\textwidth]{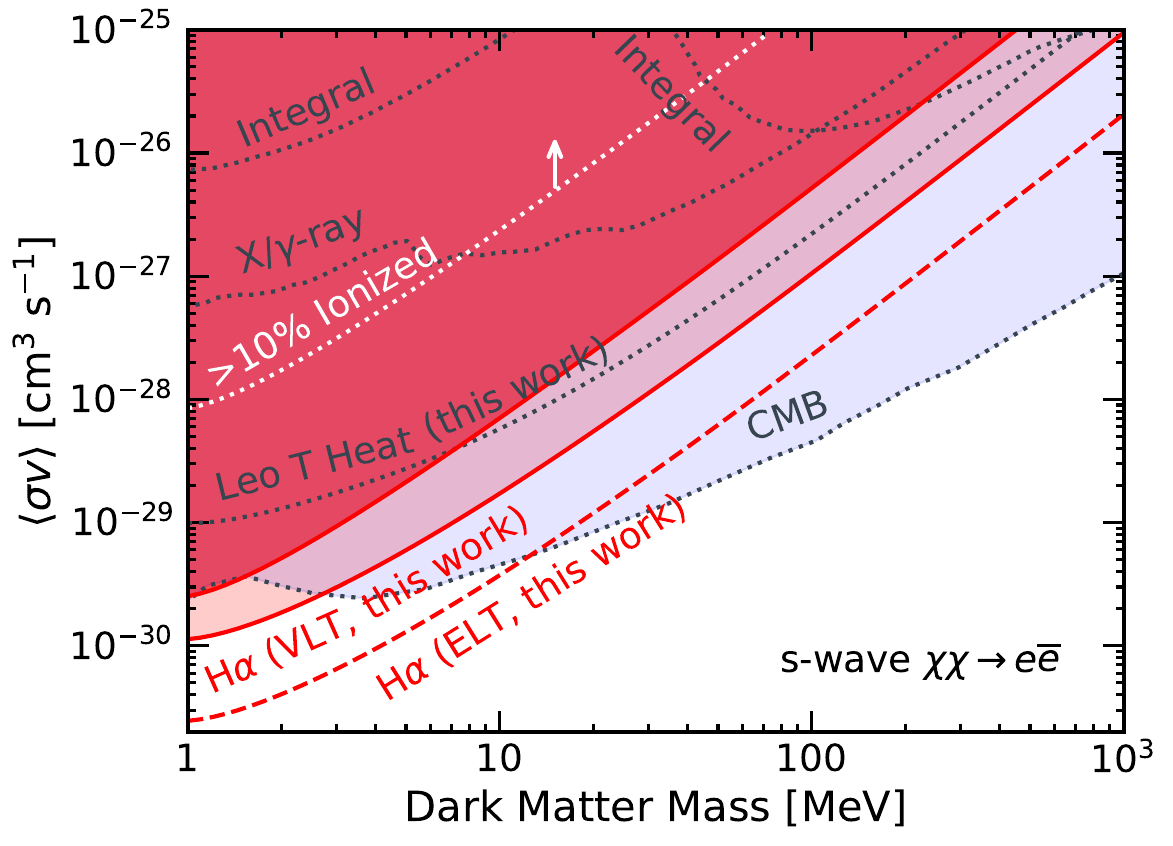}
\includegraphics[width=0.48\textwidth]{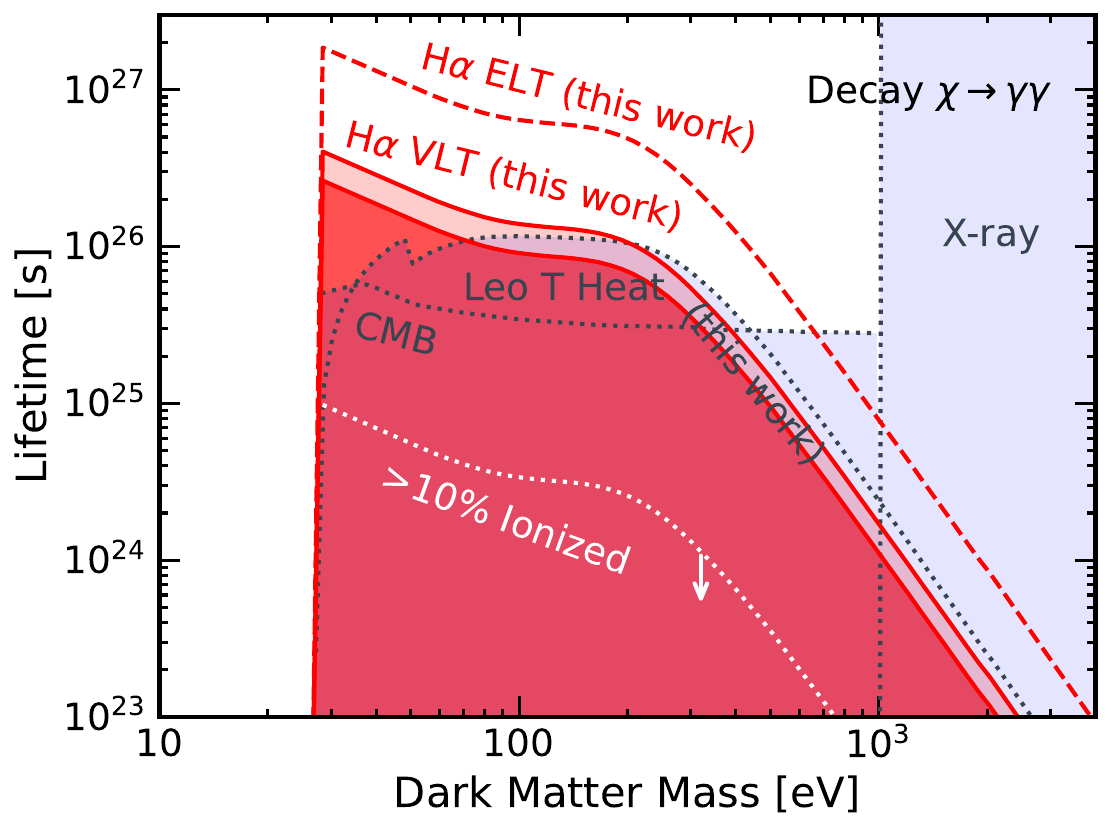}
\includegraphics[width=0.49\textwidth]{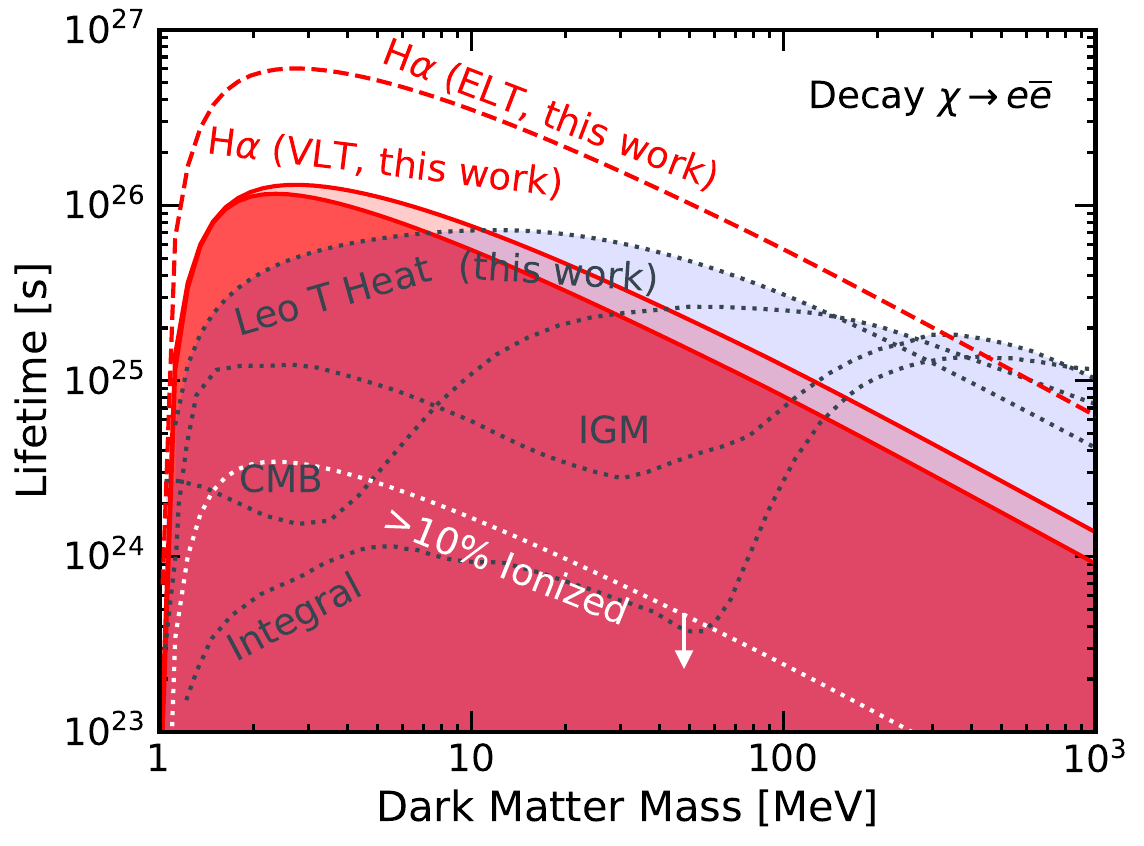}  
\vspace{-2mm}
\caption{\label{fig:results} New 95\% C.L. limits on the DM mass and annihilation rate $\langle\sigma v\rangle$ or decay lifetime into photons or electrons, using $\Ha$ measurements of the Leo T galaxy with MUSE/VLT. Dark red shaded region is for a Burkert DM profile, light red shaded is an NFW DM profile. I also show simplified projections with $\Ha$ measurements with ELT's HARMONI instrument, assuming a stack of 10 dwarfs. The white dotted line approximates where Leo T is ionized at the level of $10\%$ and therefore would no longer have a neutral medium; see text for more details. Overlaid in blue are complementary constraints from Leo T Heating (updated in this work, see also Ref.~\cite{Wadekar:2021qae}), the cosmic microwave background (CMB)~\cite{Slatyer:2015jla,Xu:2024vdn}, Integral~\cite{Cirelli:2020bpc,Cirelli:2023tnx}, a compilation of X-rays and soft $\gamma$-rays ``$X/\gamma$-ray"~\cite{Essig:2013goa}, and the intergalactic medium (IGM)~\cite{Liu:2020wqz}. These curves show the exclusion boundaries after accounting for the quoted astrophysical uncertainties where available: for the Integral constraints, the upper boundary in annihilation rate and the lower boundary in decay lifetime are shown, while for the IGM the conservative constraint is used, see text for details. Note differing axes.}
\end{figure*}

\noindent\textbf{\textit{New Constraints with the Leo T Galaxy and Future Discovery Space.---}} Leo T is an ideal target for a first application of my method. Its gas-rich neutral core and extremely low star formation/metallicity minimize conventional $\Ha$ backgrounds while providing a substantial target for ionization and recombination. Its compact neutral core and well-constrained mass profile from H\,\textsc{i} kinematics and stellar data define a clean analysis volume with a robust DM density~\cite{Ryan-Weber:2007guk,Adams_2018,Faerman:2013pmm,Patra_2018}.

The MUSE/VLT search for H$\alpha$ emission in Leo~T found no extended signal~\cite{Vaz_2023}. 
The reported 1$\sigma$  H$\alpha$ line-flux uncertainty per spaxel is 
$\sigma_{F,{\rm pix}} = 0.99\times10^{-20}\ \mathrm{erg\,s^{-1}\,cm^{-2}}$ 
for $0.2''\times 0.2''$ spaxels, $i.e.$ 
$\Omega_{\rm pix}=0.04\ \mathrm{arcsec}^2$. I combine pixels over the MUSE aperture of area $A$ using standard 
$\sqrt{N}$ statistics (with $N\equiv A/\Omega_{\rm pix}$) and include an empirically measured pixel-to-pixel covariance penalty $f_{\rm corr}$ \cite{Bacon:2014zla}.  Adopting a one-sided 95\% C.L. Gaussian multiplier $\kappa_{95}=1.645$, the aperture-integrated $\Ha$ flux limit is
\begin{equation}
F_{\Ha}^{\rm lim}
= \kappa_{95}\, f_{\rm corr}\, \sqrt{N}\, \sigma_{F,{\rm pix}}\,.
\label{eq:MUSEFluxLimit}
\end{equation}
For my field, $A=3600\ \mathrm{arcsec}^2$, 
$f_{\rm corr}=1.6$, and 
$N=A/0.04=9\times 10^{4}$, yielding $F_{\Ha}^{\rm lim} = 7.8\times 10^{-18}\ \mathrm{erg\,s^{-1}\,cm^{-2}}$. I compare the DM flux in Eq.~\eqref{eq:DMflux} to this number to set the 95\% C.L. limit. As Leo T is metal-poor, dust extinction for $\Ha$ is negligible~\cite{deJong:2008qi}; see Supplemental Material for more details.

Figure~\ref{fig:results} shows new limits on DM annihilation and decay to photons or electrons obtained with my H$\alpha$ search using MUSE observations of Leo~T. Given the uncertainty in Leo~T’s inner DM density profile, I show a band of possibilities: the darker red curve corresponds to a conservative Burkert profile, while the lighter shaded red shows a Navarro-Frenk-White (NFW) profile. The profile dependence is less pronounced for the electron case than for photons, because electrons produced outside the aperture can diffuse into the signal region, reducing sensitivity to the amount of DM in the inner core.   I also show new Leo~T heating limits, updating the treatment of heating relative to the original analysis~\cite{Wadekar:2021qae} (see also~\cite{Wadekar:2019mpc,Kim:2020ngi,Wadekar:2022ymq}); see the Supplemental Material for details. The photon heating limits show a peak-like feature just above the H and He ionization thresholds, because near threshold, the bulk of the energy goes into photoionization. This leaves little leftover kinetic energy for the photoelectrons used for temperature-based constraints. For new $p$-wave annihilation constraints, see the Supplemental Material. 

In Fig.~\ref{fig:results}, I show an illustrative projected reach for the Extremely Large Telescope (ELT) with its first-light integral-field unit (IFU) spectrograph HARMONI, assuming a stack of 10 Leo~T-like dwarfs. HARMONI is expected to achieve an observing speed gain of $\sim 25$ relative to VLT/MUSE for comparable seeing-limited IFU observations, corresponding to an idealized $\sim 5\times$ improvement in the line-flux limit at fixed exposure time~\cite{HARMONI_Performance_Oxford}. I model this by simply rescaling the MUSE flux sensitivity by this factor, effectively assuming that the relevant Leo~T H$\alpha$ emission is dominated by the central region covered by a single deep HARMONI pointing. In practice, HARMONI’s smaller field of view will modify the mapping between a given line-flux limit and the DM parameter space, especially for spatially extended signals, so the ELT curve should be viewed as a representative benchmark rather than an optimized forecast.

Fig.~\ref{fig:results} also approximates where the WNM reaches a DM-induced ionized fraction $x_e \simeq 0.1$. Beyond this point a $\Ha$ signal can still be produced, but the assumption of a weakly ionized WNM begins to break down. Exceeding this line therefore marks a complementary constraint, since such models increasingly conflict with Leo~T having an almost fully neutral H~\textsc{i} core as inferred from 21 cm observations~\cite{Ryan-Weber:2007guk,Adams_2018}.

In Fig.~\ref{fig:results}, I also show existing constraints from the CMB~\cite{Slatyer:2015jla,Xu:2024vdn}, Integral~\cite{Cirelli:2020bpc,Cirelli:2023tnx}, X-ray and gamma-ray searches~\cite{Essig:2013goa} (see also~\cite{Balaji:2025afr}), and from the IGM~\cite{Liu:2020wqz}. Where available, these curves include the quoted astrophysical uncertainties: for Integral~\cite{Cirelli:2020bpc,Cirelli:2023tnx}, I show the upper boundary in annihilation rate and the lower boundary in decay lifetime, while for the IGM~\cite{Liu:2020wqz} I use the curve labeled ``conservative.'' This avoids depicting parameter space as excluded when it remains allowed within the quantified uncertainties. Additional constraints on DM annihilation to electrons may arise from the 511~keV line, although the associated systematics are not yet under control~\cite{DelaTorreLuque:2023cef}. Voyager bounds are also omitted~\cite{Boudaud:2016mos}, since updated analyses find them weaker than other constraints over most of the relevant parameter space, with the low-mass region particularly sensitive to cosmic-ray propagation assumptions~\cite{DelaTorreLuque:2023olp}. More generally, external limits depend on assumptions about DM profiles and densities, propagation, and diffuse backgrounds, with uncertainties not always included in the quoted bounds, making direct comparisons non-trivial. By contrast, the Leo~T H$\alpha$ limits rely on a different and conservative set of astrophysical inputs, and therefore provide complementary constraints even where they appear weaker.

Compared to these existing limits, for $s$-wave annihilation to electrons, the Leo~T H$\alpha$ bounds can outperform even the leading CMB constraints for MeV-scale DM masses. The CMB limits weaken near a few MeV because injected $e^\pm$ at recombination cool mainly by inverse-Compton scattering, generating sub-Lyman photons that interact inefficiently with the gas and thus deposit little energy into ionization. For decays into photons or electrons, Leo~T also provides new leading limits over parts of parameter space; the bounds on decays to electrons in particular have important implications for model building of MeV-scale DM~\cite{Green:2017ybv}. This method probes masses below the regime where gamma-ray searches such as Fermi-LAT are most sensitive, and is complementary to early-Universe bounds, which probe a different epoch, can be suppressed or even avoided in some classes of models (see $e.g.$ Refs.~\cite{Elor:2021swj,Parikh:2023qtk}), and in general are subject to different systematics.\\


\noindent\textbf{\textit{Summary and Discovery Outlook.---}} I propose a new method to search for DM annihilation or decay using H$\alpha$ recombination emission from neutral gas. The idea is that if DM injects energy into an H\,\textsc{i}-dominated region, a predictable fraction of that power emerges as H$\alpha$ after ionization and recombination. Applying this to Leo~T with MUSE/VLT data, I obtain competitive and leading constraints across parts of the eV-GeV mass range for scenarios that inject photons or electrons. These results can be translated to specific scenarios ($e.g.$, axion-like particles, sterile neutrinos, primordial black holes); model results will appear in upcoming work. In addition, while only photons and electrons are considered here, this search can be extended to other SM states.

On the analysis side, there are clear paths to stronger sensitivity and robust discovery. Spatial DM information can be used more optimally by employing a matched filter that weights each pixel by the expected DM morphology, rather than the uniform averaging adopted here, providing an additional signal-to-noise handle. Additional recombination lines can also be incorporated, including the blue H$\beta$ line alongside red H$\alpha$, while the absence of a metal-line forest would help distinguish a DM-like recombination signal from residual star formation or shocks. Moreover, H$\alpha$ lines from DM-induced excitation, which were not included here, may further extend the reach. Beyond Leo~T, although the Milky Way Galactic Center is not optimal for Balmer lines because of strong backgrounds and dust extinction, it is plausible that, with a careful analysis, \textit{infrared} recombination lines from the higher Paschen and Brackett hydrogen series could provide an even more powerful probe there. These ideas will be explored in future work.

On the observational side, deeper integral-field or narrow-band observations of H\,\textsc{i}-dominated, DM-dominated dwarfs should push well beyond the first limits shown here. Telescopes such as Keck, MEGARA, MOONS, Magellan, and ELT offer a suite of potential improvements. DESI and Rubin/LSST will discover potentially more than 100 new Milky Way satellites, some of which may be Leo-T like~\cite{DESI:2016fyo,LSSTDESC:2025hol}. Leo-T like dwarfs can currently be seen up to about $0.5$ Mpc away; with Rubin/LSST and the Subaru Hyper Suprime-Cam the discovery reach will increase out to about $5$~Mpc~\cite{Mutlu_Pakdil_2021}. With such discoveries on the horizon, stacking multiple, similarly quiet, gas-rich dwarfs will further increase statistical power. Together, better H$\alpha$ observations and a larger sample of suitable targets can plausibly deliver orders-of-magnitude improvements in sensitivity. A discovery would appear as a narrow, spatially smooth, centrally weighted H$\alpha$ excess that is consistent across apertures and Balmer lines and lacks metal-line companions. Even \textit{without} a detection, steadily improving surface-brightness limits and multi-target analyses make H$\alpha$ a powerful new test of DM energy injection.\\

\noindent\textbf{\textit{Acknowledgments.---}} I thank Carlos Blanco, Hongwan Liu, Bernhard Mistlberger, Tracy Slatyer, Gowri Sundaresan, Daniel Vaz, and Jay Wadekar for helpful comments. RKL is supported by the U.S. Department of Energy under Contract DE-AC02-76SF00515.

\clearpage
\newpage
\onecolumngrid
\begin{center}
\textbf{\large Search for Dark Matter Annihilation and Decay with $\Ha$ Line Emission}

\vspace{0.05in}
{ \it \large Supplemental Material}\\ 
\vspace{0.05in}
{Rebecca K. Leane}
\end{center}
\onecolumngrid
\setcounter{equation}{0}
\setcounter{figure}{0}
\setcounter{section}{0}
\setcounter{table}{0}
\setcounter{page}{1}
\makeatletter
\renewcommand{\theequation}{S\arabic{equation}}
\renewcommand{\thefigure}{S\arabic{figure}}
\renewcommand{\thetable}{S\arabic{table}}

\tableofcontents

\section{P-Wave Dark Matter Annihilation Constraints}

\begin{figure*}[h]
    \centering
\includegraphics[width=0.49\textwidth]{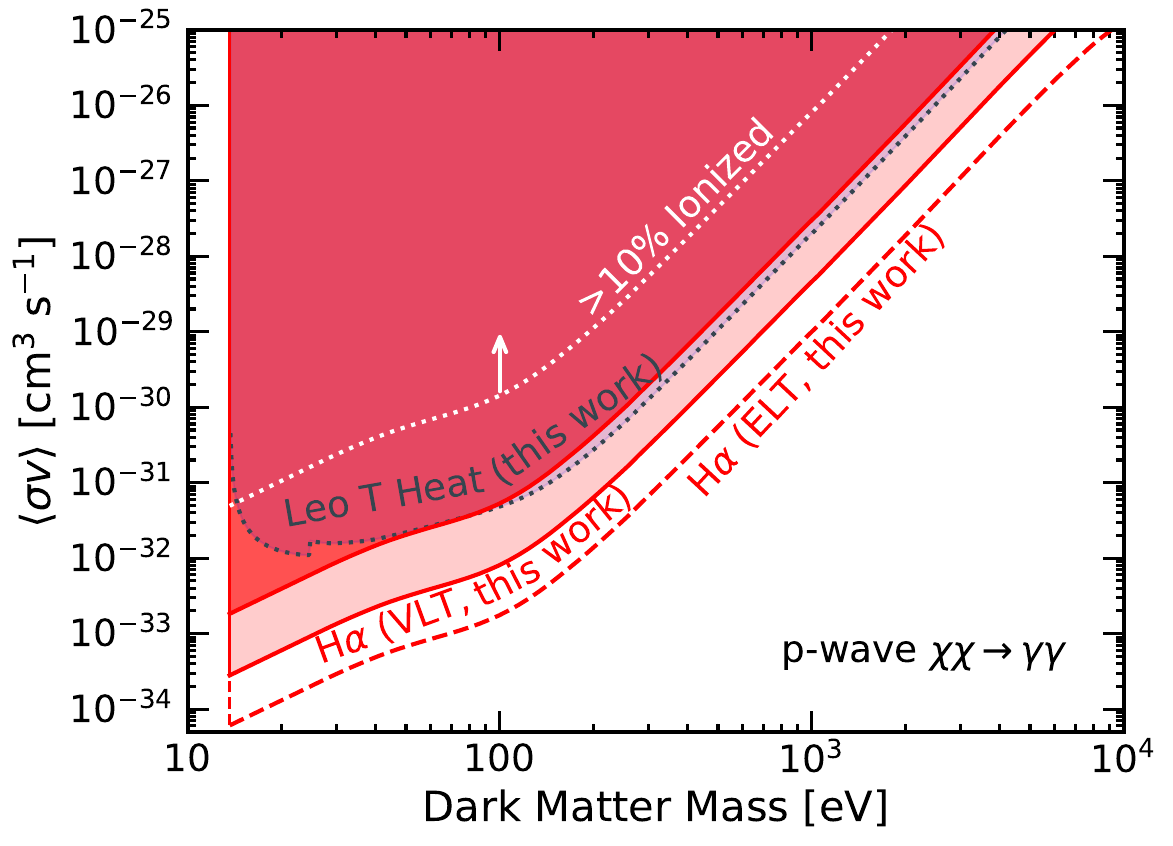}
\includegraphics[width=0.49\textwidth]{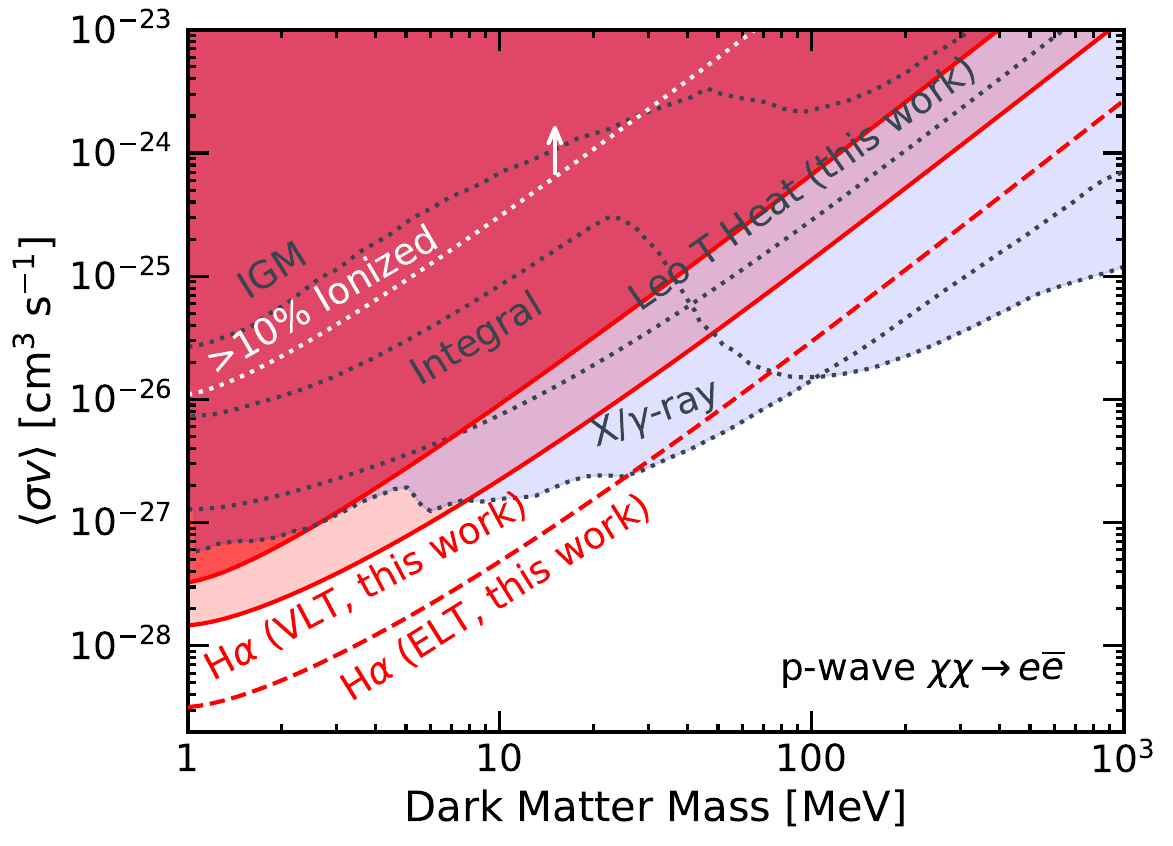}
\caption{\label{fig:resultspwave} New 95\% C.L. limits on the DM mass and $p$-wave annihilation rate $\langle\sigma v\rangle$ (evaluated at $v_{\rm ref}$) into photons or electrons, using $\Ha$ measurements of the Leo T galaxy with MUSE/VLT. Dark red shaded region is for a Burkert DM profile, light red shaded is an NFW DM profile. I also show simplified projections with $\Ha$ measurements with ELT's HARMONI instrument, assuming a stack of 10 dwarfs. The white dotted line indicates where Leo T would become ionized at the level of $10\%$ and therefore would no longer have a neutral medium; see text for more details. Overlaid in blue are complementary constraints from Leo T Heating (updated in this work, see also Ref.~\cite{Wadekar:2021qae}), Integral~\cite{Cirelli:2020bpc}, a compilation of X-rays and soft $\gamma$-rays ``$X/\gamma$-ray"~\cite{Essig:2013goa}, and the IGM~\cite{Liu:2020wqz}. Note differing axes.}
\end{figure*}

Figure~\ref{fig:resultspwave} shows $p$-wave annihilation constraints on the DM annihilation rate and mass, with new results both for $\Ha$ and heating in Leo T. To produce these constraints, I parameterize the cross section as $\langle\sigma v\rangle = \sigma_p (v/v_{\rm ref})^2$ with $v_{\rm ref} = 220~\mathrm{km/s}$, corresponding to the typical Milky Way velocity distribution. In Leo~T I take the characteristic relative speed to be $v \simeq \sqrt{6}\,\sigma_{\rm 1D}$ with $\sigma_{\rm 1D} \simeq 7.9~\mathrm{km/s}$~\cite{Wadekar:2021qae}.

\section{Leo T Gas and Dark Matter Profiles}

\begin{figure*}[h]
    \centering        \includegraphics[width=0.44\textwidth]{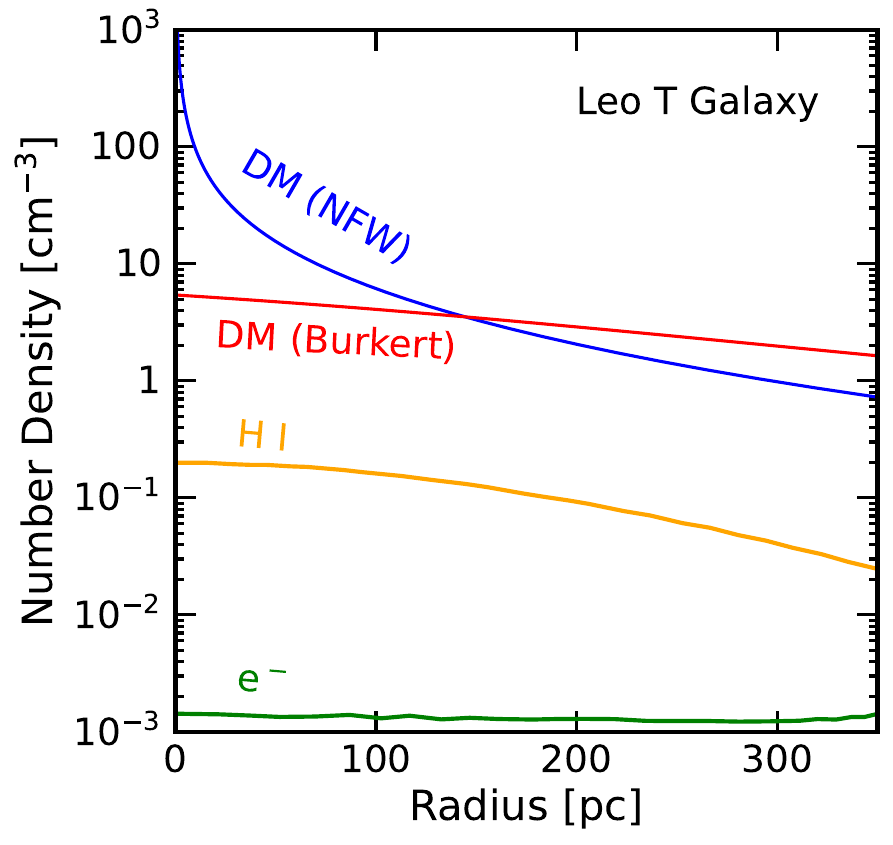} 
\caption{\label{fig:profiles} Number densities for the Leo T Galaxy. I show two DM density profiles (NFW and Burkert) for a benchmark 1 GeV DM mass, as well as the neutral gas H\,\textsc{i}, and ions $e^-$.}
\end{figure*}

Figure~\ref{fig:profiles} shows the DM and gas profiles of Leo~T; the inputs used to obtain these are discussed in the next two subsections.

\subsection{Gas}
\label{app:gas_model}

Leo~T has been studied in detail with 21\,cm facilities and deep optical spectroscopy~\cite{Faerman:2013pmm,Patra_2018,Ryan-Weber:2007guk,Adams_2018,Vaz_2023}. High-resolution Giant Metrewave Radio Telescope (GMRT) and Westerbork Synthesis Radio Telescope (WSRT) observations mapped the H\,\textsc{i} column density and kinematics, establishing Leo~T as a gas-rich, DM-dominated dwarf at $D\simeq420$\,kpc with peak H\,\textsc{i} column $N_{\rm HI}\sim7\times10^{20}\ \mathrm{cm^{-2}}$ and $M_{\rm HI}\simeq(2.8\text{–}4.1)\times10^{5}\,M_\odot$ (the larger value reflects the deeper WSRT reanalysis)~\cite{Ryan-Weber:2007guk,Adams_2018}. A Gaussian decomposition reveals both a cold neutral medium (CNM) and warm neutral medium (WNM) in the core; the CNM contains $\sim$10\% of the H\,\textsc{i} mass, while the WNM dominates the volume~\cite{Adams_2018}. The two phases have slightly offset systemic velocities, with the WNM and CNM centered at $39.6\pm0.1$ and $37.4\pm0.1\ \mathrm{km\,s^{-1}}$, respectively~\cite{Adams_2018}. MUSE spectroscopy finds no extended H$\alpha$ emission down to very low surface-brightness, consistent with the extremely low present-day star-formation activity~\cite{Vaz_2023}.

I conservatively only use the WNM, as it is better studied. For the neutral component entering my H$\alpha$ modeling, I take $n_{\rm H}(r)$ in the MUSE aperture and adopt a line-of-sight depth set by the WNM core. Using a spherical WNM core of radius $R_{\rm WNM}=350\ \mathrm{pc}$ and a uniform LOS depth of $L_{\rm depth}=700\ \mathrm{pc}$ across the MUSE field (matching the core scale, $i.e.$ from $-350$ pc to $+350$ pc), and the half-side square length of 61 pc, the resulting volume-averaged neutral-hydrogen densities are
\[
\langle n_{\rm H}\rangle_{\rm WNM\;core}\simeq 0.06\ \mathrm{cm^{-3}},\qquad
\langle n_{\rm H}\rangle_{\rm MUSE\;aperture}\simeq 0.1\ \mathrm{cm^{-3}}\,.
\]

I conservatively use averaged neutral-gas properties and a spherical WNM geometry to define the effective neutral column in the MUSE aperture, rather than attempting to model small-scale structure in the gas. As a result, the inferred $\Ha$ limits depend mainly on the effective gas column and are only weakly sensitive to modest changes in the detailed gas morphology; plausible variations shift the most gas-sensitive limits by factors of order unity rather than orders of magnitude.

A low-level ionized fraction is expected even in the predominantly neutral WNM. In the outermost region of Leo T, ionization occurs from the metagalactic UV background, while soft X-ray photons can penetrate into the inner regions~\cite{Wadekar:2021qae,Wadekar:2019mpc}. There is large uncertainty in the metagalactic UV flux, however the X-ray flux (of interest to the inner regions of Leo T) is well constrained by current observations~\cite{Haardt_2012}. In fact, the lack of detection of $\Ha$ can be used to constrain this further, though I do not study this in this work. As detailed further below, I conservatively neglect any background $e^-$ when estimating equilibrium timescales, such that the precise value of background ions is not particularly important for the results. The only other place it appears in the calculations is the ionization fraction per electron energy; taking smaller ionization fractions would only improve my results~\cite{Furlanetto_2010}.

\subsection{Dark Matter}

The DM density in Leo~T has been well studied using H\,\textsc{i} kinematics and dynamical modeling of the gas and stars~\cite{Ryan-Weber:2007guk,Adams_2018,Faerman:2013pmm,Patra_2018}. While these analyses constrain the enclosed mass profile $M(<r)$ robustly over the radii traced by the data, the exact functional form of $\rho_\chi(r)$ at arbitrary $r$ is not robustly known. Therefore, to bracket the uncertainty in the DM profile, I adopt two standard spherical parameterizations,
\begin{align}
\text{Burkert:}\qquad
\rho_\chi(r) &= \frac{\rho_0\,r_0^3}{(r+r_0)\,(r^2+r_0^2)} ,
\\
\text{NFW:}\qquad
\rho_\chi(r) &= \frac{\rho_s}{(r/r_s)\,\bigl(1+r/r_s\bigr)^2} ,
\label{eq:profiles}
\end{align}
and normalize the parameters to the dynamical constraint on the mass within $300\,$pc,
\begin{equation}
M_{300}\;\equiv\;M(<300~\mathrm{pc}) \,.
\end{equation}
From Ref.~\cite{Faerman:2013pmm}, for a Burkert profile $M_{300}\approx8\times10^{6}~$M$_\odot$, while for an NFW $M_{300}\approx6.5\times10^{6}~$M$_\odot$. I find for the Burkert profile $\rho_0=5.4$~GeV/cm$^3$, $r_0=400~$pc; for NFW $\rho_s=2.9$~GeV/cm$^3$, $r_s=350~$pc. These DM density profiles are used as inputs for Eq.~(\ref{eq:JD_region_defs}).

The dominant uncertainty in the DM profile is whether Leo~T is cored or cusped; I bracket this by showing Burkert and NFW bands in the main figures. There is additional uncertainty because only the enclosed mass at 300 pc is directly constrained, rather than the full halo parameters, but I use an aperture-averaged profile which reduces the impact of this. I have checked that my halo parameters are central values in terms of impact on the results, and that the additional uncertainty in halo parameters for the most conservative result of the Burkert profile is at the level of a few tens of percent compared to my fiducial choices. 

Leo~T’s dynamical inferences rely on H\,\textsc{i} kinematics with asymmetric-drift corrections and, where applicable, stellar velocity dispersion measurements~\cite{Ryan-Weber:2007guk,Adams_2018}. Gas pressure support, inclination, and beam-smearing systematics mainly affect the detailed shape and normalization of the rotation curve, rather than the enclosed mass at $\sim 300$\,pc where the data are best constrained. Using $M_{300}$ to normalize the halo therefore mitigates the largest modeling uncertainties.

\section{MUSE/VLT and Leo T Specific Inputs}
\label{sec:leoT_transport_inputs}

I summarize the inputs used in my study of Leo~T. The observing aperture is the Wide Field Mode of MUSE (the VLT integral-field spectrograph), a $1'\times 1'$ square. At the distance of $\sim420$ kpc this corresponds to a square of half-side $a=61$ pc centered on Leo~T. I restrict to the warm neutral medium (WNM) core of radius $R=350$ pc, so the line-of-sight (LOS) depth across the core is $L_{\rm depth}\simeq 700$ pc. 

\subsection{Electrons}

For electrons, as detailed in the main text, I consider the case where electrons can be injected into the neutral gas and wander in or out of the aperture from surrounding regions. I denote the aperture column as zone $A$ and the complementary WNM volume as zone $B$. Exterior to the WNM core is zone $C$; to be conservative zone $C$ is considered as an exit region but not an input region, as it is no longer the neutral medium.

For the $A\to B$ channel, I use the rectangular-column eigenmode with absorbing sides,
\begin{equation}
\ell_{A\to B} \;=\; \frac{a}{\sqrt{2}} \simeq 43 ~\mathrm{pc}.
\end{equation}
$A\to C$ is neglected as it is comparably a negligible escape route.

For the $B$ born channels I compute DM-weighted effective lengths from the harmonic mean of the local gap distances. Gas weighting is conservatively neglected; including $n_H(r)$ would modestly decrease $L_{B\to A}$ and increase $L_{B\to C}$ because the gas peaks toward the core, strengthening the limits. The resulting lengths are:

\begin{itemize}
    \item Leaving Region B to Region A: For NFW profile, $\ell_{B\rightarrow A}=45$ pc (annihilation) and $\ell_{B\rightarrow A}=60$ pc (decay). For the Burkert profile, $\ell_{B\rightarrow A}=60$ pc (annihilation) and $\ell_{B\rightarrow A}=70$ pc (decay).
    \item Leaving Region B to Region C:  For NFW profile,  $\ell_{B\rightarrow C}=175$ pc (annihilation) and $\ell_{B\rightarrow C}=125$ pc (decay). For the Burkert profile,  $\ell_{B\rightarrow C}=115$ pc (annihilation) and $\ell_{B\rightarrow C}=100$ pc (decay).
\end{itemize}

In Eq.~(\ref{eq:diffusion}) of the main text I adopt the ion density $n_{\rm ion}(r)$ from Fig.~\ref{fig:profiles} and, conservatively, a uniform magnetic field $B=1~\mu$G throughout. I take a coherence length of $L_{\rm coh}=50\ \mathrm{pc}$; it is often even smaller in dwarf galaxies ($\sim1-20$ pc)~\cite{1996ApJ...458..194M}, so this should be conservative.

\subsection{Photons}

Compared to electrons, the photon scenario is much more straightforward, as I conservatively assume ballistic, isotropic emission. To compute the optical depth I use the aperture-averaged gas density within the MUSE column and adopt a mean photon path length of $\ell_{\rm gas}\simeq 101\ \mathrm{pc}$. This is the mean exit distance for rays launched from the center of a rectangular box with half-sizes $(61,61,350)\ \mathrm{pc}$ ($i.e.$, the $122\times122\times700\ \mathrm{pc}$ MUSE aperture). This is a good approximation as the DM profiles and gas densities are weighted towards the center. I have checked that profile-specific weighting changes the results less than the $\sim10$ percent level.

\subsection{Attenuation of the \texorpdfstring{$\Ha$}{Halpha} signal}

The emergent H$\alpha$ signal is negligibly attenuated both within Leo~T and along the line of sight to the Milky Way. Within Leo~T, attenuation would require a non-negligible population of hydrogen atoms in the $n{=}2$ state to resonantly absorb H$\alpha$ photons. In the predominantly neutral WNM, however, this population is tiny. A conservative estimate is obtained by taking the number density of hydrogen atoms in the $n{=}2$ state, $n_2$, to be set by the recombination rate times the lifetime of the longer-lived $2s$ state. For a fiducial background free-electron density $n_e^{\rm bg}\sim 10^{-3}\,\mathrm{cm^{-3}}$ and case-B recombination coefficient $\alpha_B\simeq2.6\times10^{-13}\,\mathrm{cm^3\,s^{-1}}$, the recombination rate per unit volume is $\alpha_B (n_e^{\rm bg})^2\sim 3\times10^{-19}\,\mathrm{cm^{-3}\,s^{-1}}$. Taking the $2s$ lifetime to be $\sim 0.1$~s gives
\begin{equation}
   n_2\sim \alpha_B (n_e^{\rm bg})^2 \times 0.1~{\rm s}\sim 3\times10^{-20}\,\mathrm{cm^{-3}}. 
\end{equation}
Over a path length of order $L\sim700$ pc, this corresponds to a column density of $n{=}2$ hydrogen atoms of
\begin{equation}
N_2\sim n_2 L\sim 60\,\mathrm{cm^{-2}}.  
\end{equation}
Using the standard Doppler-broadened line-center absorption cross section for the $n{=}2\to3$ transition at $T\sim7000$~K gives $\sigma_0({\rm H}\alpha)\sim 6\times10^{-13}\,\mathrm{cm^2}$. Multiplied with the column density $N_2$ above, this then gives an H$\alpha$ optical depth of order $\tau_{\mathrm{H}\alpha}\sim10^{-11}$. Thus self-absorption of H$\alpha$ within Leo~T is completely negligible.

Along the line of sight from Leo~T to the Milky Way, attenuation at optical wavelengths is also negligible. A conservative upper bound can be obtained from Thomson scattering by free electrons, for which the optical depth is approximately $\tau_T\sim n_e \sigma_T L$, where here $n_e$ is the free-electron density along the line of sight, $\sigma_T$ is the Thomson cross section, and $L$ is the propagation distance. Even taking $n_e\sim10^{-4}\,\mathrm{cm^{-3}}$ over the full $\sim420$ kpc path gives only $\tau_T\sim10^{-4}$ (the two electron densities refer to different environments: $n_e^{\rm bg}\sim10^{-3}\,\mathrm{cm^{-3}}$ is a fiducial local background value inside the Leo~T WNM, whereas $n_e\sim10^{-4}\,\mathrm{cm^{-3}}$ is a conservative characteristic density for the much more diffuse gas along the external line of sight to the Milky Way). 

The only relevant correction is therefore foreground Milky-Way dust extinction. For Leo~T, the foreground reddening is modest, $E(B-V)\simeq0.03$--$0.04$~\cite{deJong:2008qi}, corresponding to $A_{\mathrm{H}\alpha}\simeq0.08$--$0.1$ mag, $i.e.$ only a $\sim7$--$10\%$ attenuation in flux. This is subdominant to the other astrophysical uncertainties, so I do not include an additional correction in the limits.

\section{Comparison with Leo T Heating}

The use of Leo T for DM heating signatures was first pointed out in Ref.~\cite{Wadekar:2019mpc} in the context of DM scattering (see also Refs.~\cite{Chivukula:1989cc,Bhoonah:2018wmw,Bhoonah:2020dzs,Lu:2020bmd,Kim:2020ngi,Bhoonah:2018gjb} for DM-induced ISM heating in general), and applied to DM annihilation and decay in Ref.~\cite{Wadekar:2021qae}, producing strong constraints. Here I make some improvements to the calculation for DM annihilation and decay resulting in detectable Leo T heating; the results for these improved constraints are shown in the main text. To produce the updated constraints I follow the methodology of Ref.~\cite{Wadekar:2021qae}, but with some improvements that I now detail below.

First, I self-consistently include the effects of primordial helium in the gas. Helium was previously included in the opacity, but not the threshold effects relevant for helium's photoionization. Specifically, I include the fact that if DM annihilates or decays into photons, and the energy of the photons are above helium's ionization threshold of 24.6 eV, then helium can be photoionized, which reduces the energy available for the electron ejected by helium, which then reduces the amount of heating. This is the uptick/peak observed near helium's ionization threshold in Fig. 2 of the main text. Note there is an additional uptick also shown near hydrogen's ionization threshold; this was included already in Ref.~\cite{Wadekar:2021qae}.

\begin{figure*}[t]
    \centering        \includegraphics[width=0.42\textwidth]{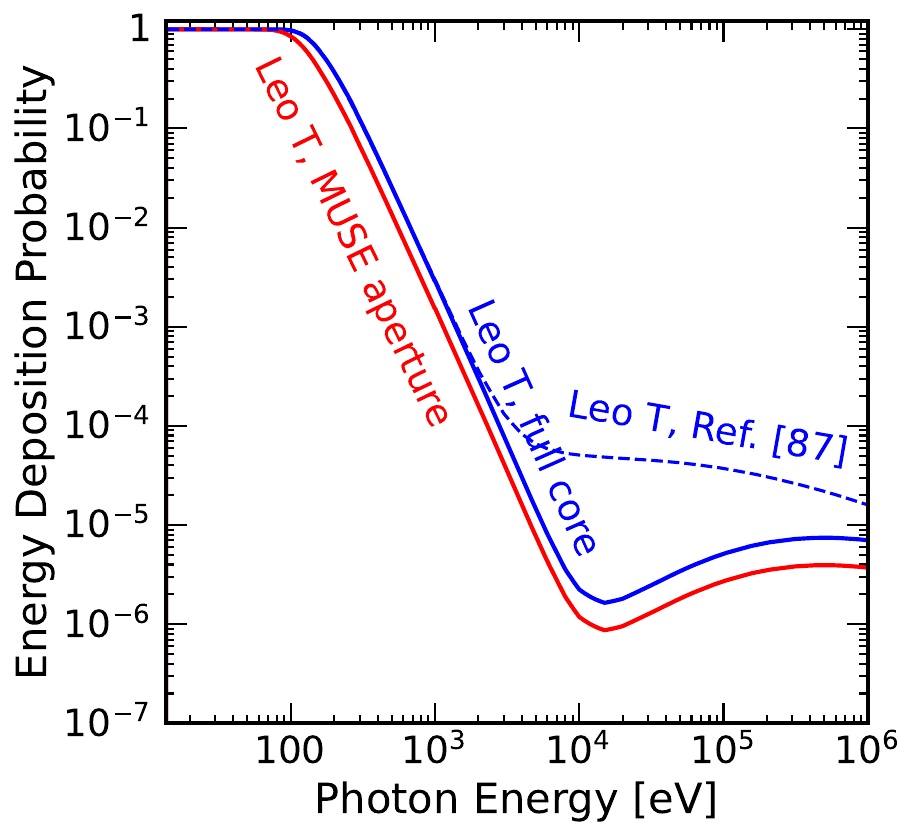}
    \includegraphics[width=0.42\textwidth]{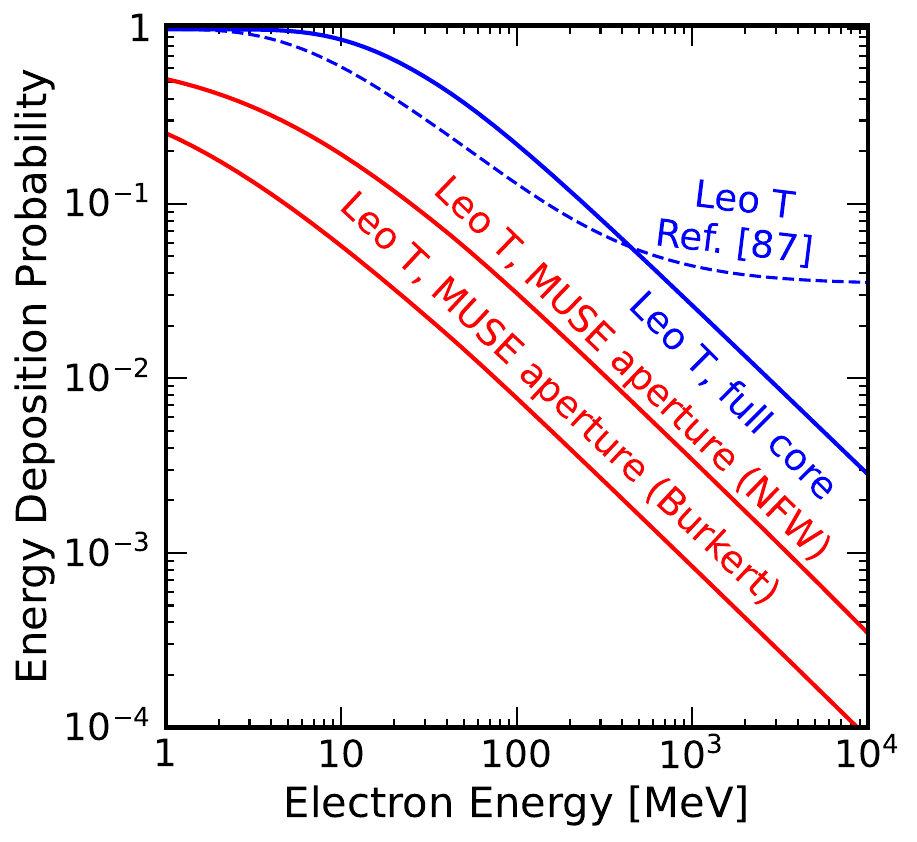}
\caption{\label{fig:energy} Energy deposition probability for photons and electrons in Leo T. Comparison is shown for the MUSE aperture relevant for $\Ha$ limits (red solid), as well as if the whole WNM core is used (blue solid). I also show for comparison the result following the setup from Ref.~\cite{Wadekar:2021qae}, which assumes the full WNM relevant for DM heating, but uses the total
attenuation / stopping rates as if they correspond to local energy deposition (plus some other differences); see text for details.}
\end{figure*}

Second, for photons I recompute the energy-deposition probability using the mass
energy-absorption coefficient, which removes the contribution from coherent Rayleigh scattering and, for Compton/Thomson scattering, counts only the fraction of the photon energy that is actually transferred to electrons. In the low-column Leo~T environment, scattering mainly redirects photons rather than fully thermalizing them locally. Ref.~\cite{Wadekar:2021qae} instead used the total attenuation coefficient, effectively treating all interactions as true absorption, which overestimates the deposited fraction once scattering begins to compete with photoelectric absorption at higher energies. The impact of this change is visible in the photon energy deposition curves at higher energies, in Fig.~\ref{fig:energy} (left).

\begin{figure*}[t]
    \centering        \includegraphics[width=0.47\textwidth]{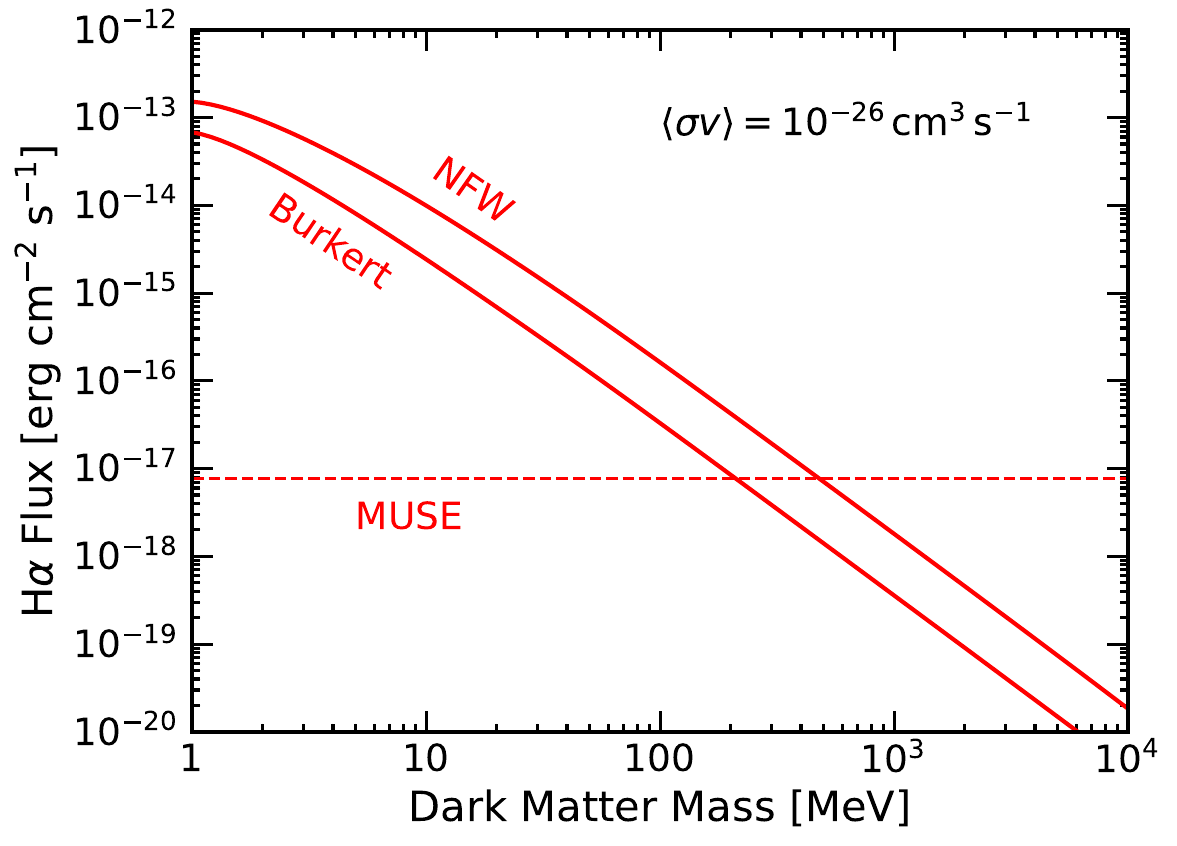}
    \includegraphics[width=0.47\textwidth]{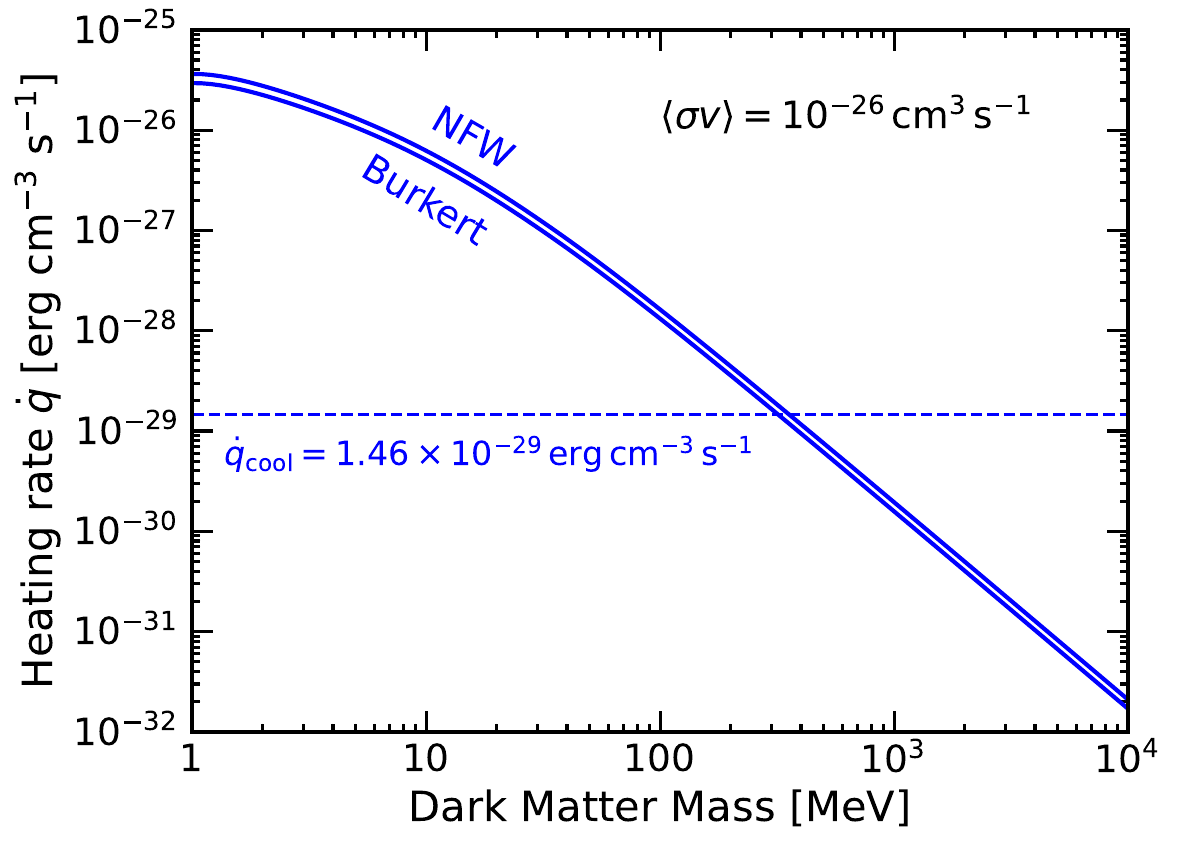}
\caption{\label{fig:heatingvshalpha}Comparison of fluxes and limits for both $\Ha$ and heating probes with Leo T, for an example DM annihilation cross section $\langle\sigma v \rangle=10^{-26}\,$cm$^3$/s, assuming annihilation into electrons, and assuming either an NFW or Burkert DM profile. Dashed lines are the experimental measurements used for exclusions.}
\end{figure*}

Third, I update the treatment of the stopping power for electrons. Ref.~\cite{Wadekar:2021qae} included both collisional and radiative (bremsstrahlung) stopping power when determining the amount of energy deposited into the gas. However, only the collisional component leads to appreciable local heating and ionization in Leo T; the bremsstrahlung photons produced at these energies traverse the neutral core with optical depths $\tau \ll 1$ and therefore escape without depositing a significant fraction of their energy. In this work, I therefore keep only the collisional stopping power when computing the deposition probability for electrons. This weakens the heating results at higher electron energies where radiative losses begin to dominate, $i.e.$ above $\sim 80$~MeV. The difference can be seen by comparing the blue solid line in Fig.~\ref{fig:energy} (right), which uses only the collisional stopping power, with the blue dashed line, which follows the total stopping-power prescription of Ref.~\cite{Wadekar:2021qae}.  Note Fig.~\ref{fig:energy} is for the example case of DM annihilation.

Fourth, for the fraction of electron energy that is ultimately deposited as heat or ionization, I use the Monte Carlo results of Ref.~\cite{Furlanetto_2010} instead of the older analytic fit of Ref.~\cite{Ricotti:2001zf} adopted in Ref.~\cite{Wadekar:2021qae}. Ref.~\cite{Furlanetto_2010} argues that a single analytic form cannot accurately describe all heating/ionization fractions, and recommends using simulation-based fits for the system-specific ionization fractions; I therefore employ their heating fractions appropriate for a primordial H/He gas. In the high-energy limit these simulations give heating fractions that are $\sim 20\%$ smaller than the analytic fit, leading to correspondingly $\sim 20\%$ weaker heating constraints. This is a modest change but I include it here for completeness.

Finally, both the $\Ha$ limits and the heating limits require the diffusion coherence length, which is relevant for the electron diffusion behavior. In this work I use $L_{\rm coh}=50$ pc, which is conservative given the value is often smaller ($\sim1-20$ pc) in dwarf galaxies~\cite{1996ApJ...458..194M}. Ref.~\cite{Wadekar:2021qae} was exceptionally conservative and took 100 pc; for fair direct comparison I have recomputed the heating bounds also with the choice of 50 pc. In Fig.~\ref{fig:energy} (right, electron panel), this is why the solid line is higher than the dashed blue (which instead used 100 pc coherence lengths) up to electron energies of about 200 MeV; choosing instead 100 pc would lower the solid blue by about a factor of 2 in the optically thin regime. This is not necessarily an improvement but rather a less conservative input choice.

Figure~\ref{fig:heatingvshalpha} shows the difference in the detectability of DM signals for $\Ha$ and heating, using the $\Ha$ signal from this work as well as the updated heating bound from this work, compared to their experimental observables (measured diffuse $\Ha$ flux vs cooling rates). For a benchmark DM annihilation cross section of $10^{-26}\,{\rm cm}^3/{\rm s}$ and $m_\chi = 1~{\rm MeV}$, the expected $\Ha$ signal lies about four orders of magnitude above the detection threshold, while the corresponding DM heating signal lies about three orders above. This explains the roughly order-of-magnitude difference between the resulting constraints in the optically thick limit: with MUSE, the $\Ha$ line is intrinsically more sensitive than the cooling-rate probe. However, at higher DM masses we enter the optically thin regime: the medium is no longer fully absorbing, so the fraction of energy deposited depends on the traversed column density, and the total signal scales with the size of the region of interest. This trend is also visible in Fig.~\ref{fig:energy}, where at higher energies in the optically thin regime there is a larger difference between using the full WNM core and only the smaller MUSE aperture. Because the cooling constraint integrates over the full WNM core while the $\Ha$ bound only uses the MUSE field of view, the smaller region of interest (ROI) for $\Ha$ eventually weakens its constraints relative to heating. A future instrument that mapped the entire core in $\Ha$ would therefore strengthen the $\Ha$ limits, whereas the heating bounds would not improve comparably. The behavior with different DM profiles follows the same logic. For $\Ha$, the sensitivity improves for an NFW profile relative to a Burkert profile, because the MUSE aperture is more concentrated on the inner region where NFW is denser. For heating, by contrast, the signal integrates over the entire core, so the central enhancement is less important, and the difference between NFW and Burkert is much smaller.

In Fig.~\ref{fig:energy}, for the electron case, NFW clearly yields a larger in-aperture signal than Burkert, because the DM density is more concentrated within the MUSE field of view.
For photons, I do not show separate Burkert and NFW energy-deposition curves. This is not because the profiles are irrelevant, but because they only enter through the $J$ or $D$ factors, which are applied after the energy-deposition calculation. The curves in Fig.~\ref{fig:energy} therefore show only the probability that a photon deposits its energy in region A; to obtain the total flux, these probabilities are later multiplied by the appropriate in-aperture DM density (the relevant $J$ or $D$ factor for the MUSE aperture for H$\alpha$, or for the full WNM core for heating).
For electrons, in contrast, the in-aperture deposition already depends on how DM-produced $e^\pm$ move between regions. Electrons born in region B can random-walk into region A, and the rate at which this happens depends directly on the DM profile through the relative weighting of A and B. As a result, in the electron case the energy-deposition fractions for both the H$\alpha$ and heating probes are combined with the whole-core $J$/$D$ factors at this stage, while the photon case only involves region A and the profile dependence is folded in later. This is why the electron panel in Fig.~\ref{fig:energy} already shows a larger apparent difference between the H$\alpha$ and heating curves, whereas for photons the full difference only appears once the smaller MUSE-aperture $J$/$D$ factors (for H$\alpha$) are contrasted with the larger whole-core factors (for heating).

Note that in Fig.~\ref{fig:energy} (left panel), the electrons produced consequently from the photon ionizations are assumed to be absorbed in the aperture region with full efficiency. Given the low energies of interest for the photon cases, this is a valid approximation. In the case where the DM energy and therefore the secondary electron energy approaches the MeV scale, there will be some suppression that increases with energy (see right panel). I do not enter this regime for the main text's photon results.

\section{Dark Matter Ionization and Hydrogen Recombination Equilibrium} 

As DM ionizes the medium it creates free electrons, while hydrogen recombinations remove them. The free-electron density $n_e$ therefore evolves as
\begin{equation}
  \frac{dn_e}{dt} \;=\; {\zeta_\chi\,n_{\rm H}}-\alpha_B(T)\,n_e^2,
  \label{eq:evolve}
\end{equation}
where $\zeta_\chi$ is the DM-induced ionization rate per H atom, $n_{\rm H}$ is the neutral hydrogen density, and $\alpha_B(T)$ is the case-B recombination coefficient,
\begin{equation}
  \alpha_B(T) \;\equiv\; \sum_{n\ge 2}\alpha_n(T) \;=\; \alpha_A(T)-\alpha_{n=1}(T)\,,
\end{equation}
with  $\alpha_A(T)$ the Case~A recombination coefficient, which is instead recombination in the optically thin regime. In Case~B, recombinations directly to the ground state ($n{=}1$) emit Lyman-continuum photons whose mean free path in neutral gas is negligible; these photons are absorbed on the spot and re-ionize a nearby atom, so they do not change the net free-electron density. For neutral hydrogen at temperatures between about 4,000 to 16,000 K, a good approximation is \cite{1989agna.book.....O,2022MNRAS.510.2797P,RydenPogge2021_IISM}
\begin{equation}
  \alpha_B(T) \simeq 2.6\times 10^{-13}
  \left(\frac{T}{10^4\,\mathrm{K}}\right)^{-0.7}\ \mathrm{cm^3\,s^{-1}}.
\end{equation}
The fraction of Case-B recombinations that produce an H$\alpha$ photon is
\begin{equation}
  f_{H\alpha}(T) \;\equiv\; \frac{\alpha^{\rm eff}_{H\alpha}(T)}{\alpha_B(T)}
  \;\approx\; 0.46 \quad \text{(for }T\sim 7000\,\mathrm{K}\text{)}\,.
\end{equation}
From Eq.~\eqref{eq:evolve}, the equilibrium timescale is
\begin{equation}
    t_{\rm eq}=\frac{1}{\sqrt{\alpha_B\,\zeta_\chi\,n_H}}\,.
\end{equation}
To take into account whether the processes are in equilibrium or not, I use an equilibrium fraction,
\begin{equation}
  f_{\rm eq} \;=\; \tanh^2\!\left(\frac{t_{\rm av}}{t_{\rm eq}}\right)\in(0,1],
  \label{eq:feq}
\end{equation}
where $t_{\rm av}$ is the effective duration over which local conditions are steady ($e.g.$, the quiescent time), which for Leo T I take as $\sim50$ Myr. In the steady-state limit $t_{\rm av}\!\gg\!t_{\rm eq}$, $f_{\rm eq}\!\to\!1$; for $t_{\rm av}\!\ll\!t_{\rm eq}$ one has $f_{\rm eq}\simeq (t_{\rm av}/t_{\rm eq})^2$. Here I have conservatively assumed zero additional background electrons; their inclusion would only speed up the process. Including a small ambient free-charge density ($e.g.$, $n_e^{\rm bg}\sim10^{-3}\,\mathrm{cm}^{-3}$) would increase equilibration and improve limits up to an order of magnitude; I have conservatively neglected this.

To make direct connection to the variables used in the main text, Eq.~\eqref{eq:feq} can also be written as
\begin{equation}
f_{\rm eq}
=\tanh^2\!\Big[t_{\rm av}\,\sqrt{\alpha_B\,\zeta_\chi\,n_{\rm H}}\;\Big]\,,
\label{eq:feq_def}
\end{equation}
where $\alpha_B$ is the case-B recombination coefficient, $n_{\rm H}$ the neutral hydrogen density, $t_{\rm av}$ the effective accumulation timescale, and $\zeta_\chi=\overline{\mathcal{Q}}[A] \,f_{\rm dep}^{\Ha}(E) /(n_{\rm H}f_{\Ha}E_{\Ha})$ is the DM-induced ionization rate per H atom, with $\overline{\mathcal{Q}}[A]=\mathcal{Q}[A]/(\Omega_A\,L_{\rm depth})$, where $L_{\rm depth}$ is the depth of the ROI (in the case of Leo T, as detailed above,  $L_{\rm depth}\simeq 700$ pc).

\bibliography{main}
\bibliographystyle{apsrev4-2}

\end{document}